\documentclass[%
 aip,
 amsmath,amssymb,
preprint,%
]{revtex4-1}

\usepackage{url}
\usepackage{graphicx}
\usepackage{dcolumn}
\usepackage{bm}
\usepackage{braket}

\usepackage[utf8]{inputenc}
\usepackage[T1]{fontenc}
\usepackage{mathptmx}
\usepackage{etoolbox}
\usepackage[dvipsnames]{xcolor}
\usepackage{soul}

\makeatletter
\newcommand{\thickhline}{%
    \noalign {\ifnum 0=`}\fi \hrule height 1pt
    \futurelet \reserved@a \@xhline
}
\newcolumntype{"}{@{\hskip\tabcolsep\vrule width 1pt\hskip\tabcolsep}}
\makeatother

\makeatletter
\def\@email#1#2{%
 \endgroup
 \patchcmd{\titleblock@produce}
  {\frontmatter@RRAPformat}
  {\frontmatter@RRAPformat{\produce@RRAP{*#1\href{mailto:#2}{#2}}}\frontmatter@RRAPformat}
  {}{}
}%
\makeatother

\newcommand{\angstrom}{\mbox{\normalfont\AA}}

\let\hl\undefined
\newcommand{\hl}[1]{#1}

\begin{document}

\preprint{AIP/123-QED}

\title[Complexity Reduction in Density Functional Theory: Locality in Space and Energy]{Complexity Reduction in Density Functional Theory: Locality in Space and Energy}
\author{William Dawson}
\email{william.dawson@riken.jp}
\author{Eisuke Kawashima}%
\affiliation{ 
RIKEN Center for Computational Science, Kobe, Hyogo,
650-0047, Japan.
}%
\author{Laura~E.~Ratcliff}
\affiliation{Centre for Computational Chemistry,
School of Chemistry, University of Bristol, Bristol BS8 1TS,
United Kingdom.}
\author{Muneaki Kamiya}
\affiliation{Faculty of Regional Studies, Gifu University, Gifu, 501-1132 Japan}
\author{Luigi Genovese}%
\affiliation{Univ.\ Grenoble Alpes, CEA, IRIG-MEM-L\_Sim, 38000 Grenoble, France.}
\author{Takahito Nakajima}%
\affiliation{ 
RIKEN Center for Computational Science, Kobe, Hyogo,
650-0047, Japan.
}%

\date{\today}

\begin{abstract}
We present recent developments of the NTChem program for performing large scale hybrid Density Functional Theory calculations on the supercomputer Fugaku. We combine these developments with our recently proposed Complexity Reduction Framework to assess the impact of basis set and functional choice on its measures of fragment quality and interaction. We further exploit the all electron representation to study system fragmentation in various energy envelopes. Building off this analysis, we propose two algorithms for computing the orbital energies of the Kohn-Sham Hamiltonian. We demonstrate these algorithms can efficiently be applied to systems composed of thousands of atoms and as an analysis tool that reveals the origin of spectral properties.
\end{abstract}

\maketitle

\section{Introduction}
Kohn-Sham Density Functional Theory (KS-DFT)~\cite{hohenberg-inhomogeneous-1964, kohn-self_consistent-1965} is a commonly used framework for performing quantum mechanical simulations of materials and molecules. One of the drawbacks of using KS-DFT is that the standard algorithms have a computational cost that grows with the third power of the system size, significantly limiting the class of systems that can be tractably studied. Walter Kohn's nearsightedness principle~\cite{PhysRevLett.76.3168} offers one possible solution to this problem. According to this principle, the elements of the one particle density matrix decay exponentially with distance in systems with a finite gap (and metals at high temperature). This property has been used to develop new algorithms for KS-DFT calculations which have a computational cost which grow only linearly with the system size~\cite{bowler-O(N)-2012, ratcliff-2016-challenges}. \hl{For the study of molecular systems using Gaussian orbitals, this locality has been successfully employed for many years to develop linear scaling algorithms for computing exact-exchange for hybrid functionals}~\cite{Burant1996,Schwegler1996,Schwegler1997,Ochsenfeld1998,Kussmann2013}.

The nearsightedness principle is not only useful for computational purposes, but can also form the basis of analysis techniques of large systems~\cite{Dawson2022}. In a recent series of papers~\cite{Mohr2017,10.1021/acs.jctc.9b01152}, we have proposed a "Complexity Reduction Framework" (hereafter called QM-CR), which exploits this principle to partition systems into well defined fragments and quantify inter-fragment interactions. QM-CR was implemented in the BigDFT program~\cite{doi:10.1063/5.0004792} and has proven useful for understanding substrate-ligand binding~\cite{Chan2021} as well as the role of mutations in protein-protein binding~\cite{zaccaria2022probing}.

One limitation of our previous works has come from the particularities of the BigDFT basis set. In BigDFT, the KS orbitals are represented in a basis of in-situ optimized support functions that are in turn represented in an underlying basis of wavelets. This basis set is unusual, making it unclear how transferable the framework is to other KS-DFT programs. The basis set is also optimized for representing only the occupied orbitals in the pseudopotential approximation, leaving the role of core electrons and unoccupied states unclear. BigDFT is further limited to calculations using semi-local functionals for large systems. 

NTChem~\cite{Nakajima2015} is a quantum chemistry program based on Gaussian basis sets. In recent years, we have implemented a low order scaling algorithm in NTChem for performing hybrid DFT based on the PrelinK screening method~\cite{Kussmann2013}. This implementation is aimed at computing large systems using the supercomputer Fugaku. Here we will describe recent improvements in NTChem, which allow us to utilize QM-CR to post-process large scale hybrid DFT calculations. We will focus on the problem of computing the eigenvalues of the KS Hamiltonian and show how this process can be accelerated by exploiting locality in space and energy.

\section{Complexity Reduction Overview}

We begin by summarizing QM-CR. In KS-DFT, we represent the KS orbitals in some set of $M$ basis functions:
\begin{equation}
\psi_i = \sum_j^M c_{ij}\phi_j.
\end{equation}
This leads to basis set representations of fundamental operators:
\begin{align}
S_{ij} = \bra{\phi_i}\hat{I}\ket{\phi_j} \\
H_{ij} = \bra{\phi_i}\hat{H}\ket{\phi_j} \\
K_{ij} = \bra{\phi_i}\hat{K}\ket{\phi_j}
\end{align}
where $\hat{I}$ is the identity operator, $\hat{H}$ the KS Hamiltonian, and $\hat{K} = \sum_i^{M} f_{i} \ket{\psi_i} \bra{\psi_i}$ the density operator (with occupation numbers $f_i$). These definitions lead to the KS eigenvalue problem:
\begin{equation}
H\psi_i = \epsilon_i S \psi_i.
\end{equation}
The L{\"o}wdin representation of the density matrix $\ddot{K} = S^{\frac{1}{2}}KS^{\frac{1}{2}}$ (Hamiltonian $\ddot{H} = S^{-\frac{1}{2}}HS^{-\frac{1}{2}}$), is a projection operator on to the subspace defined by the occupied orbitals. Restricting the occupation numbers $f_i$ to either $1$ or $0$ leads to the following properties:
\begin{align}
\mathrm{Tr}(\ddot{K}) = N \\
\ddot{K} = \ddot{K}\times \ddot{K} - \ddot{K}.
\end{align}

We note that while the density matrix is usually defined using occupation numbers according to the aufbau principle, this is not the only projection that can be constructed. Much like the density matrix defined by the occupied orbitals, any projection we construct will be sparse if there exists a gap between the eigenvalues of the orbitals included/excluded from the projection~\cite{Benzi2013}. For this study, we will  define $\ddot{K}^{C}$ as the projection on to only the core orbitals. For $\ddot{K}$ the sparsity is defined by the HOMO--LUMO gap and for $\ddot{K}^C$ the (often significantly larger) core--valence gap.

\subsection{Purity and Bond Order}

Suppose that we partition the basis functions into two sets representing \emph{fragments} of the system $A$ and $B$. We can then define sub-blocks of the matrix as being associated with either a given fragment (block-diagonal) or fragment pair (block-off-diagonal). We introduce two measures of these blocks:
\begin{equation}
q_{AA}\Pi_{AA} = \mathrm{Tr}(\ddot{K}_{AA}\ddot{K}_{AA} - \ddot{K}_{AA})
\end{equation}
where $q_{AA}$ is a normalization factor equal to the total number of electrons of the isolated fragment in the gas phase and:
\begin{equation}
B_{AB} = \mathrm{Tr}(\ddot{K}_{AB} \times \ddot{K}_{BA}).
\end{equation}
We call $\Pi$ the \emph{purity indicator} and interpret it as a descriptor of fragment quality. This is because it measures the degree of idempotency of the block $\ddot{K}_{AA}$ and thus its quality as a projection. We have previously proposed that an (absolute) purity value of $0.05$ be used as a cutoff for determining whether a system fragmentation is reasonable or not. $B$ is called the \emph{fragment bond order} and measures the chemical interaction between a pair of fragments. It measures the off diagonal terms that are present in the block matrix--matrix multiplication calculation of the idempotency condition of the full matrix which are ignored if we treat the system as being block diagonal (non-interacting fragments). Our previous results have shown that $B$ can be used to automatically construct embedding environments for a target fragment. This is done by including fragments until the sum of the bond order between the target fragment and all excluded fragments is below $0.001$. 

\subsection{\hl{Fragment Projected Density of States}}
\label{subsec:pdos}

Once a fragmentation has been determined, a block of $\ddot{K}$ can be used to project eigenvectors of $\ddot{H}$ on to the occupied subspace of a given fragment. This is similar to what is often done to compute the Projected Density of States (DoS), where the projection is done on to orbitals associated with different atom types and angular momentums~\cite{10.1103/PhysRevB.16.3572,Sanchez-Portal1995,aarons2019atom}. Following the analysis in our previous work~\cite{Mohr2017,10.1021/acs.jctc.9b01152}, we here generalize this process to projection on to arbitrary fragments. 


Consider again a system composed of two fragments A and B. Much like we can apply $\ddot{K}$ to a matrix of column vectors $V$ to project them on the occupied subspace of $\ddot{H}$, we can also apply the sub-block $\ddot{K}_{AA}$ to the upper part of the matrix $\big[ V_{AA}|V_{AB} \big]$ (diagonal block $AA$ and off--diagonal $AB$). The suitability of this operation depends on whether $\ddot{K}_{AA}$ can be treated as a projection matrix (i.e. the purity of fragment $A$). In the case where $V$ is the eigenvectors of $\ddot{H}$, we can write:
\begin{equation}
W = \big[ V_{AA}|V_{AB} \big]^T \ddot{K}_{AA} \big[ V_{AA}|V_{AB} \big]
\end{equation}
where the diagonal of the matrix $W$ contains weights which describe the overlap of the original vectors and the projected ones. These weights can be used to assign eigenvectors and eigenvalues to a given fragment.

\section{NTChem Developments}

In this section, we will present specific developments of the NTChem program which will enable this study. The implementation of large scale hybrid DFT was first presented in a Japanese language article~\cite{Nekir}, so we reproduce key details here.

\subsection{Large Scale Hybrid Density Functional Theory}

In NTChem, evaluation of two-electron integrals is done analytically using the SMASH program~\cite{Ishimura2008} and is accelerated by using the PreLinK screening scheme developed by Ochsenfeld and coworkers~\cite{Kussmann2013, Kussmann2015}. When performing analytic integrals on large systems with a sparse density matrix, the cost of screening integrals (which grows like $O(N^4)$) can become a bottleneck, particularly when density matrix elements are further reduced in magnitude through the \emph{difference densities} technique~\cite{Haser1989}. The PrelinkK method addresses this bottleneck by computing bounds on elements of the Hamiltonian matrix. If an element falls below some threshold, the calculation can be skipped. The cost can further be reduced by block sorting integrals such that once an integral is successfully screened, all subsequent evaluations are skipped. We use a similar procedure for computing the Coulomb contribution.

One of the challenges of performing hybrid DFT calculations on large systems is that the density matrix is too large to replicate in memory across all processes. The memory problem is particularly acute on a machine like Fugaku which only has 32GB of memory per node shared between 48 cores. In recent years this bottleneck has been overcome by shared memory implementations~\cite{Ishimura2010, mironov2017efficient, mironov2019efficient}; \hl{on supercomputers with a large amount of memory per node (e.g. 256GB on the ARCHER2 supercomputer) this can greatly expand the size of systems that can be computed, though even these implementations eventually have limits}. Several codes have gone further by employing distributed matrix datastructures~\cite{Foster1996, Furlani2000, Alexeev2002, Umeda2010, Liu2014, Chow2016, NWChem, Huang2020}\hl{, which in addition to increasing the available memory avoid the communication cost of a global reduction operation}. In NTChem, we distribute the Hamiltonian and density matrices using the distributed sparse matrix datastructure implemented in NTPoly~\cite{DAWSON2018154}. NTPoly uses a three-dimensional data distribution where the matrix is partitioned along the $X$ and $Y$ axes and replicated in $Z$. This allows for a reduction in the memory use, while offering communication savings when extra memory is available. \hl{This distribution is also used for the exchange and correlation evaluation. Processes which hold the same matrix elements (same $X$ and $Y$ coordinate but different $Z$) store and evaluate the density and potential on different grid points.}

Once the Hamiltonian has been constructed, it is necessary to compute the density matrix. In standard DFT implementations, the density matrix is computed through solving the eigenvalue problem. Such an approach is available in NTChem using the EigenExa library~\cite{imamura2011development}. Alternatively, a diagonalization free approach based on matrix functions can be employed as implemented in NTPoly. In this study we will use the fourth order trace resetting method~\cite{Niklasson2003}. In NTPoly, sparsity of the matrix is maintained by filtering values below a certain threshold to zero. This leads to an efficient implementation of sparse matrix--sparse matrix multiplication that can be parallelized using the same three dimensional distribution described earlier.

\subsection{Python Framework for Driving Calculations}
\label{sec:example}

To facilitate the complex workflows required for simulating large systems, we have modified the PyBigDFT~\cite{doi:10.1063/5.0004792} Python framework for use with NTChem. Beyond the creation of input files and calculators for NTChem, we have also developed a basis set class that automatically fetches information from the Basis Set Exchange using its RestAPI~\cite{Pritchard2019}. Input parameters can be setup in Python and are written automatically as a Fortran namelist for NTChem to read. To reduce the cost of the SCF cycle, we construct a fragment based guess using the adjustable density matrix assembler method (ADMA)~\cite{Mezey2014}, with hydrogen capping done automatically based on the system's connectivity matrix. Fragment charges can be assigned from fragment types or through interoperability with OpenBabel's various charge models~\cite{OBoyle2011}.

\begin{figure}
    \centering
    \includegraphics[width=0.8\columnwidth]{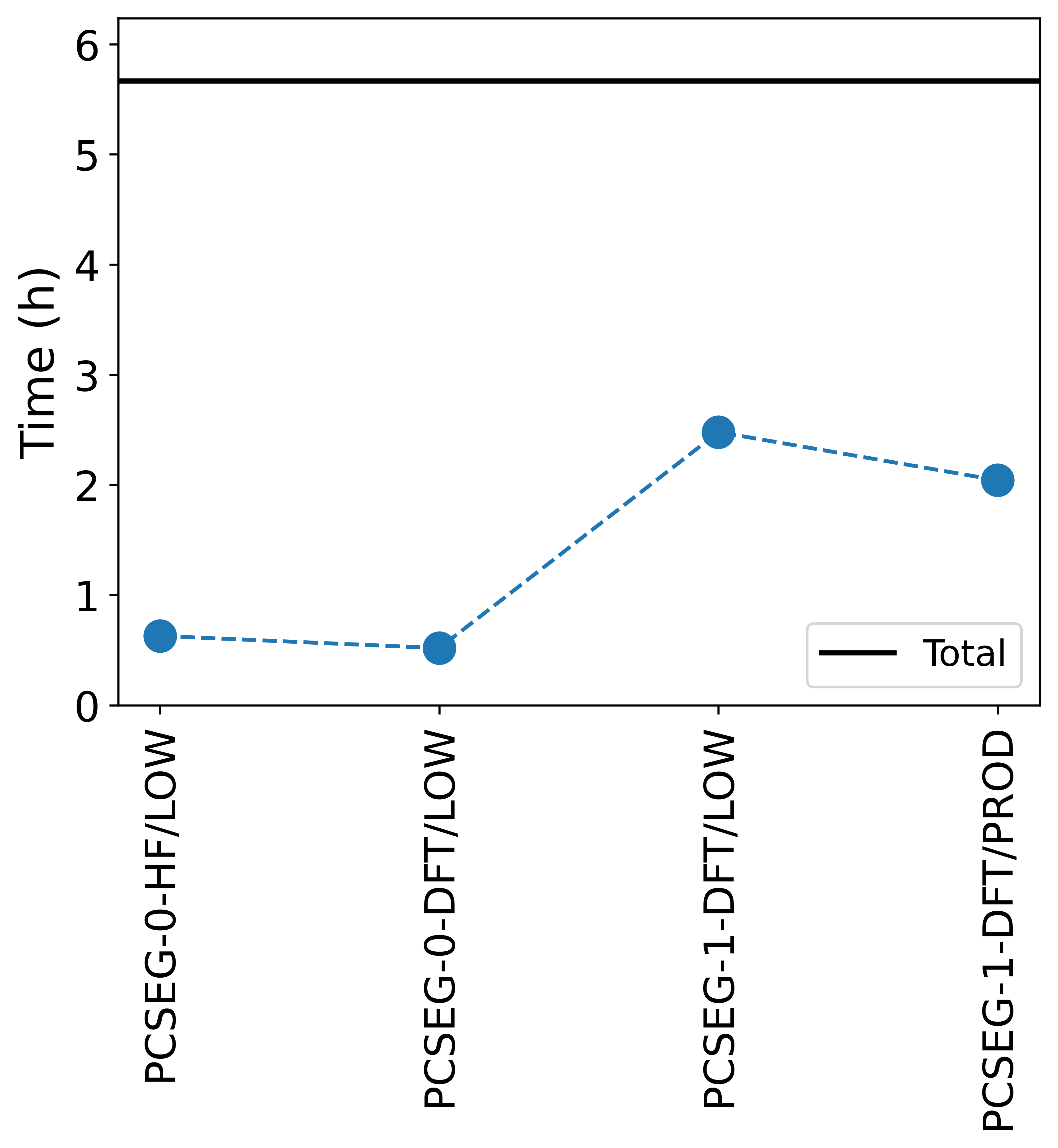}
    \caption{\hl{Calculation} time required for a converged hybrid DFT calculation  of the 1CRN protein in a NaCl solution using 1024 nodes of Fugaku. \hl{Each node had four MPI processes and 12 OpenMP threads.} With the largest basis set there are 8539 atoms and 69413 basis functions. Each calculation requires only a few SCF iterations using the CDIIS technique~\cite{Pulay1982}, with the PCSEG-1-DFT/LOW requiring the most at 14 iterations.}
    \label{fig:timingdft}
\end{figure}

As an example of this combined framework, we consider the calculation of a small protein (PDB: 1CRN~\cite{Teeter1984}) in a salt water solution (Figure~\ref{fig:timingdft}) \hl{computed using the free boundary condition}. The system was generated using pdbfixer and equilibrated using the Amber99sb forcefield~\cite{Wang2000,Hornak2006} as implemented in OpenMM~\cite{eastman2017openmm} (see Supplementary Information Sec.~I for a visualization). The PyBigDFT framework partitions the systems into fragments made up of either protein amino acids, ions, or clusters of water molecules. We utilize the Polarization Consistent basis sets series \hl{(PCSEG)}~\cite{Jensen2014} because it has been shown that PCSEG-0 serves as an excellent guess for projecting to larger basis sets~\cite{Lehtola2019}. 

The first set of calculations are done with Hartree--Fock (HF), for accelerated convergence, and subsequently with the BHandHLYP~\cite{Becke1993} (DFT) functional. The LOW accuracy parameter set uses the SG-1 grid~\cite{Gill1993}, a Schwartz screening threshold of $1\times 10^{-9}$, a PrelinK screening of $1\times 10^{-7}$ for Coulomb and $1\times 10^{-4}$ for exchange, a level-shifting parameter of 0.5 Hartree, and a total energy convergence criteria of $1\times 10^{-4}$ Hartree. For the \hl{production (PROD)} parameters we use an increased grid size \hl{($N^r=99$, $N^\Omega=590$)}, a Schwartz screening threshold of $1\times 10^{-12}$, a PrelinK screening of $1\times 10^{-8}$ for Coulomb and $1\times 10^{-5}$ for exchange, no level shifting, and a total energy convergence criteria of $1\times 10^{-5}$ Hartree. \hl{In the final iteration, the change in the density matrix $||K^{OUT} - K^{IN}||_2$ is below $1\times 10^{-8}$.}

These calculations were performed on 1024 nodes of Fugaku. We note that the supercomputer Fugaku has 158,976 nodes, and the "large" system queue on Fugaku begins at 385 nodes, such that this can be viewed as a reasonable amount of computational resources for any Fugaku project. The entire workflow is encapsulated in a Jupyter notebook which can be run on the front end of the Fugaku machine and can launch jobs using the scheduler.

\section{Eigenvalue Calculations Local in Space and Energy}

One of the drawbacks of the calculation of the density matrix using diagonalization free approaches is that the eigenvalues of the Hamiltonian are useful for system analysis. For example, they would be useful for computing the DoS using Koopmans' theorem \hl{(i.e. by approximating the excitation energies as the negative of the Kohn-Sham eigenvalues)}. While in general the accuracy of a Koopmans' approach is low, it can be improved significantly through the use of advanced functionals or correction schemes~\cite{Leeuwen1994,Casida2000,Baer2010,Tsuneda2010,Gritsenko2016,Teale2022}. Recently, our group has investigated the use of range separated hybrid functionals~\cite{Hirao2020, Hirao2021b, Hirao2022} and correction schemes based on mixing of several calculations~\cite{Hirao2021, Chan2021b}. The DoS of just the core orbitals can be useful for interpreting X-ray photoelectron spectroscopy~\cite{Norman2018}. With this in mind, there would be significant value to a method that could reliably compute those values without fully diagonalizing the Hamiltonian. However, as the KS orbitals are not local in space, and must be kept orthogonal to one and other, it is challenging to derive low order scaling approaches. A number of methods have have been proposed that exploit information gained during the purification process~\cite{Mohr2017-b, Kruchinina2018} to approximate select eigenvalues. Another family of approaches is based on the kernel polynomial method~\cite{Lin2016}, which approximates the DoS through random sampling. Select eigenvalues might also be computed using a shift-and-invert approach~\cite{10.1145/1089014.1089019,10.1145/3409571,10.1145/3404397.3404416,Lee2018} or contour integration~\cite{10.1103/PhysRevB.79.115112, Nakata2017}. Dense eigenvalue solvers also can solve for only part of the spectrum (for example, doing only a partial back transformation in the ELPA 2 algorithm~\cite{Marek2014}, or using subspace filtering as implemented in the ChASE library~\cite{10.1145/3313828}).

\begin{figure}
    \centering
    \includegraphics[width=1\columnwidth]{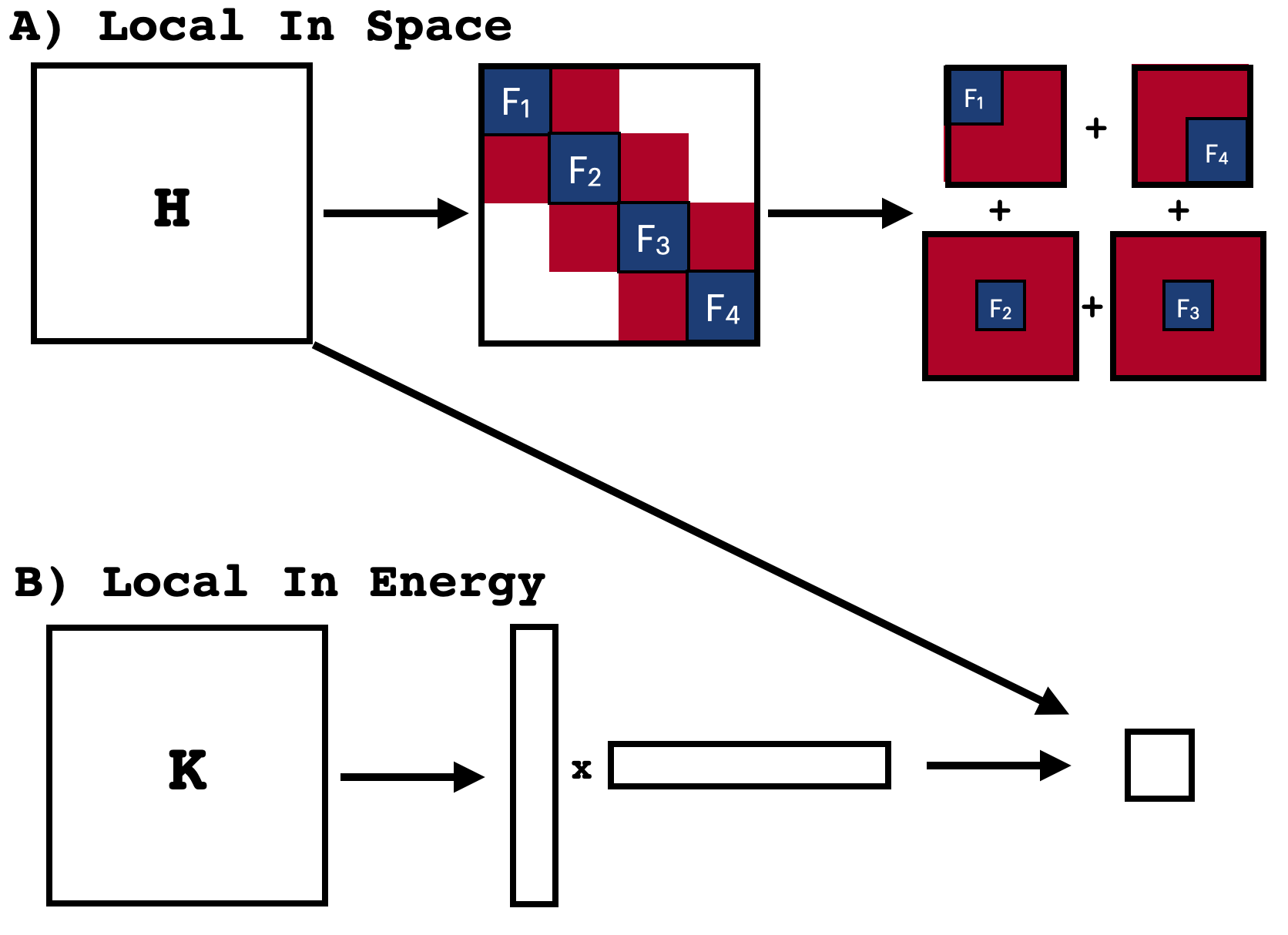}
    \caption{Diagram of the two eigenvalue algorithms proposed in this work: (A) local in space and (B) local in energy.}
    \label{fig:diagram}
\end{figure}

The first solution we propose is to exploit locality in space (Figure~\ref{fig:diagram}A). By joining up fragments of a system automatically using $\Pi$ as a guide, we can derive a partitioning of the system for which the Hamiltonian can be approximated as block diagonal. The eigenvalues of individual blocks can then be computed and combined together for the whole spectrum. The limitation of this approach will be atoms that sit on the boundary between fragments. To remedy this, we can compute a given fragment in a buffer defined by $B$, and project the result on to the original fragment (Sec.~\ref{subsec:pdos}). When sufficiently pure fragments are used, the weights will unambiguously assign a state to a given fragment, avoiding the risk of double counting. In our previous work, we compared the DoS computed from a full system calculation and summed up from independent fragment calculations using different $\Pi$ cutoffs~\cite{10.1021/acs.jctc.9b01152}. This analysis was done as validation of our fragmentation procedure and found that the DoS could be qualitatively reproduced using a $\Pi$ value of $0.01$. In practice, we already have the Hamiltonian of the full system, so there is no need to run independent calculations. By directly operating on the total Hamiltonian, we anticipate being able to compute a more accurate approximation. 

We also propose to exploit locality in energy (Figure~\ref{fig:diagram}B). Since $\ddot{K}$ is symmetric, positive semidefinite, it can be decomposed using the pivoted Cholesky decomposition into a set of $N$ orthogonal orbitals $\ddot{K} = LL^T$, where $N$ is the rank of $\ddot{K}$. Like the density matrix, the Cholesky vectors are sparse, and can thus be computed efficiently~\cite{Aquilante2006}. These orbitals can then be used to perform a similarity transformation of the Fock matrix $\ddot{H}^{\mathbf{Occ}} = L\ddot{H}L^T$ reducing the problem to diagonalizing a smaller $N \times N$ matrix. This approach is inspired by the divide and conquer eigenvalue algorithm of Nakatsukasa and Higham~\cite{Nakatsukasa2013}. In our case, we are able to exploit sparsity by replacing the QR decomposition with pivoted Cholesky and using density matrix purification to construct the projection. This approach is also similar to the spectrum splitting algorithm proposed by Motamarri et al.~\cite{10.1103/PhysRevB.95.035111} which is tailored towards the matrix free approach using a finite element basis.

We note that with both these algorithms it is in principle possible to retrieve as well an approximation to the eigenvectors. For the local in space algorithm, the eigenvectors of the small matrix would simply be padded with zeros. This will lead to a set of vectors which are not quite orthogonal, but the accuracy may be sufficient for analysis techniques. For the local in energy algorithm, the eigenvectors of the full matrix are the product of the Cholesky vectors and the eigenvectors of $\ddot{H}^{\mathbf{Occ}}$. The accuracy of these vectors will depend on the threshold used for for filtering small matrix values.

\section{Basis Set, Functionals, Energy Envelopes}

We now will utilize QM-CR as implemented with NTChem to analyze the role of locality in space and energy. We will first look at the impact of the choice of basis set and functional in comparison with the original implementation in the BigDFT code. We will then look at QM-CR values when evaluated with projections associated to different energy windows.

\subsection{Impact of Basis Set and Functionals}

\begin{figure*}
    \centering
    \includegraphics[width=0.85\textwidth]{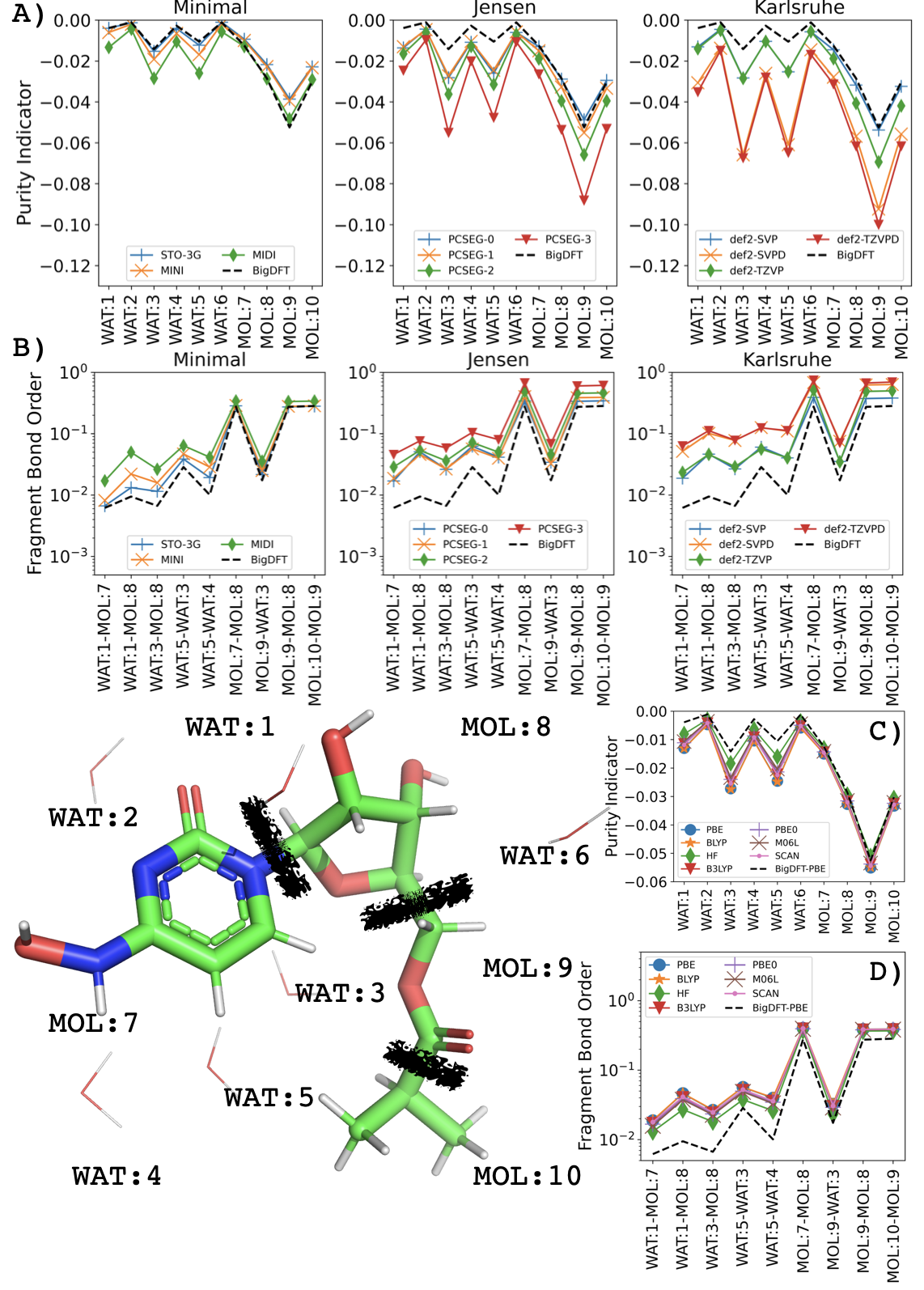}
    \caption{Comparison of QM-CR quantities computed with BigDFT and NTChem using different basis sets and functionals. The dependence of $\Pi$ (A) and $B$ (B) on the choice of basis set. The dependence of $\Pi$ (C) and $B$ (D) on the choice of functional \hl{(computed with the PCSEG-1 basis set)}. A 3D representation of Molnupiravir is shown, with each of its fragment labeled (image generated by PyMOL~\cite{PyMOL}).}
    \label{fig:cgau}
\end{figure*}

We begin by examining the impact of the choice of basis set and functional on the QM-CR quantities (Figure~\ref{fig:cgau}). As a choice of system, we investigate a Molnupiravir molecule surrounded by water molecules generated using Packmol~\cite{Martinez2009} and then relaxed with the GAFF forcefield~\cite{Wang2004} as implemented in OpenBabel. This system is small enough to examine in detail, while being large enough to partition into fragments (done in this case using chemical intuition). We study three classes of basis sets: Minimal (STO-3G and Huzinaga's MINI and MIDI~\cite{huzinaga2012gaussian}), Jensen (PCSEG-0, PCSEG-1, PCSEG-2, PCSEG-3~\cite{Jensen2014}), and Karlsruhe (def2-SVP, def2-SVPD, def2-TZVP, def2-TZVPD~\cite{Weigend2005, Rappoport2010}). For functionals, we examine generalized gradient approximation (GGA) functionals (PBE~\cite{10.1103/PhysRevLett.77.3865} and BLYP~\cite{10.1103/PhysRevA.38.3098, 10.1103/PhysRevB.37.785}), hybrid functionals (B3LYP~\cite{Stephens1994} using VWN5 correlation~\cite{vosko1980accurate}, PBE0~\cite{Adamo1999}), Hartree--Fock (HF), and two meta-GGAs (M06L~\cite{Zhao2006} and SCAN~\cite{PhysRevLett.115.036402}). BigDFT calculations were performed in the linear scaling mode using a wavelet grid spacing of $0.37$ atomic units, HGH pseudopotentials~\cite{Willand2013}, and the PBE functional.

We use the QM-CR values computed with BigDFT as a reference for comparison. BigDFT's quasi-orthogonal, in-situ optimized basis functions are similar to Wannier functions, and are thus a more phyiscal representation of the system's inherent sparsity. By contrast, when using Gaussian orbitals, the use of higher angular momentum or more diffuse functions makes it less justified to assign a given basis function to a given atom. By comparing between the two representations, we can see the impact of basis set on QM-CR quantities.

Our analysis finds (Figure~\ref{fig:cgau}A-B) that larger basis sets have less pure fragments and stronger inter-fragment interactions (higher fragment bond order). The diffuse functions of the Karlsruhe basis set family lead to even worse purity values than the quadruple-$\zeta$ quality PCSEG-3 basis set (with similar fragment bond order values). The BigDFT values are most similar to the minimal basis sets for the water molecules, yet are more consistent with the double-$\zeta$ quality representations of the Molnupiravir molecule. This shows the inherent challenge of designing basis sets that are both local and able to represent a complex chemical environment, without in-situ optimization. The compactness of the orbitals for the water molecules suggests that there may be benefit in tailoring a basis set for the representation of atoms in a water molecule (one focused on the liquid phase~\cite{Schutt2018}, as opposed to a more general set~\cite{Corsetti2013}).

Nonetheless, the overall trends remain quite similar, even across basis set families. This is also true when comparing different functionals (Figure~\ref{fig:cgau}C-D). GGA functionals are less localized than those that include exact exchange, as well as the meta-GGAs M06L and SCAN. The HF result stands out as the most compact representation. This matches our experience of using the ADMA guess where its convergence is more efficient with HF than DFT. This locality may be physically justified, given the impressive accuracy of density corrected DFT~\cite{Sim2022}. In Supplementary Information Sec.~II we show a more detailed analysis of the effect of varying the fraction of exact exchange. We conclude that most functionals capture well the fundamental coarse-grained structure of this simple system (though it is known that they can differ substantially if observed in more detail~\cite{Medvedev2017}). Increasing the cardinality of the basis set or adding diffuse functions can lead to interpretability challenges, similar to known deficiencies of L{\"o}wdin charge analysis. However, since large fragments are used, as guided by the purity indicator, QM-CR appears to be useful even for Gaussian basis set codes if cutoffs are reconsidered and care is taken about the impact of basis set and functional.

\subsection{Further Functional Validation}

\begin{figure}
    \centering
    \includegraphics[width=1\columnwidth]{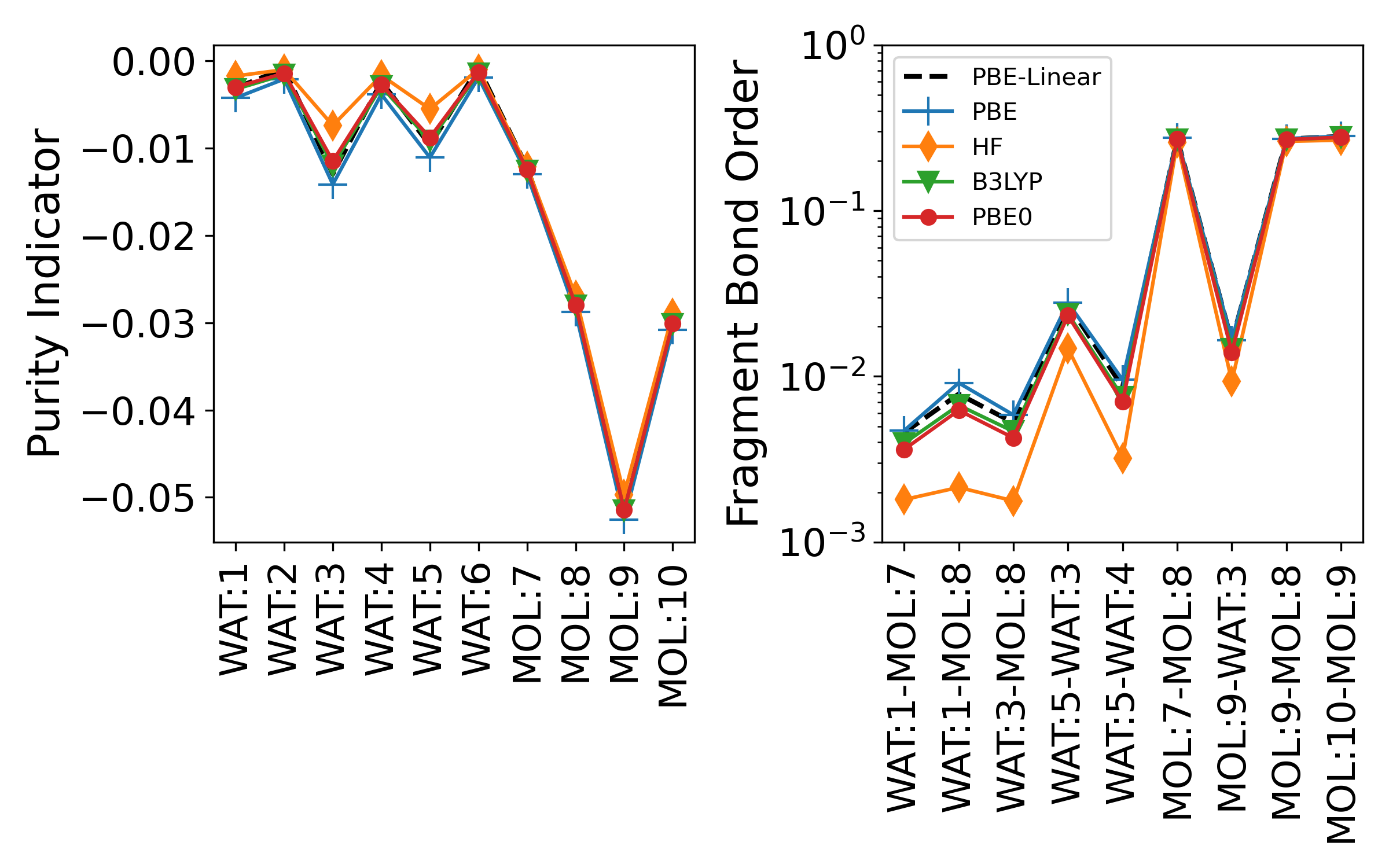}
    \caption{Comparison of $\Pi$ values and $B$
    computed using different functionals fitted to the PBE optimized BigDFT basis set. "PBE-Linear" is the fully converged linear scaling calculation at the PBE level. All other results are from fitted cubic scaling calculations.}
    \label{fig:linear-cubic}
\end{figure}

BigDFT can be run in two different modes: cubic and linear. In the cubic mode, the KS orbitals are directly represented in the wavelet basis set. In the linear scaling mode, the KS orbitals are represented by the in-situ optimized support functions. While hybrid functionals have not been implemented for the linear scaling version, such calculations are possible in the cubic scaling mode~\cite{Ratcliff2018}. As further validation of our analysis of the effect of functional, we propose the following workflow. First, we will compute the optimized basis functions using a linear scaling run of BigDFT with the PBE functional. Then we will perform a cubic scaling calculation with PBE and various hybrid functionals. The KS orbitals from the cubic scaling calculation will then be projected to the optimized basis functions, which will be used to construct the density matrix in the basis of the PBE optimized support functions, from which we can compute the QM-CR quantities (Figure~\ref{fig:linear-cubic}).

We see a very similar pattern as that which emerged from fully converged calculations in the Gaussian basis sets. The inclusion of exact exchange increases the purity of the fragments and decreases the fragment bond order. The fitting process itself is not perfect. If we compute the trace of $K^{F}S$ where $K^{F}$ is the fitted density matrix from the cubic PBE calculation and $S$ the overlap matrix from the linear scaling calculation with PBE, the error introduced is $0.11$ electrons. The error in energy, $\mathrm{Tr}(HK^F) - \mathrm{Tr}(HK)$ using $H$ from the linear calculation, is $0.035$ Hartree (relative error of 0.04\%). However, the QM-CR quantities are quite similar when comparing the fully converged PBE linear scaling calculation and the fitted PBE. From these results, we conclude that integration of QM-CR with the NTChem code is a promising strategy for understanding the influence of more sophisticated functionals on system partitioning and inter-fragment interactions.

\subsection{Core Density Matrices}

\begin{figure}
    \centering
    \includegraphics[width=1\columnwidth]{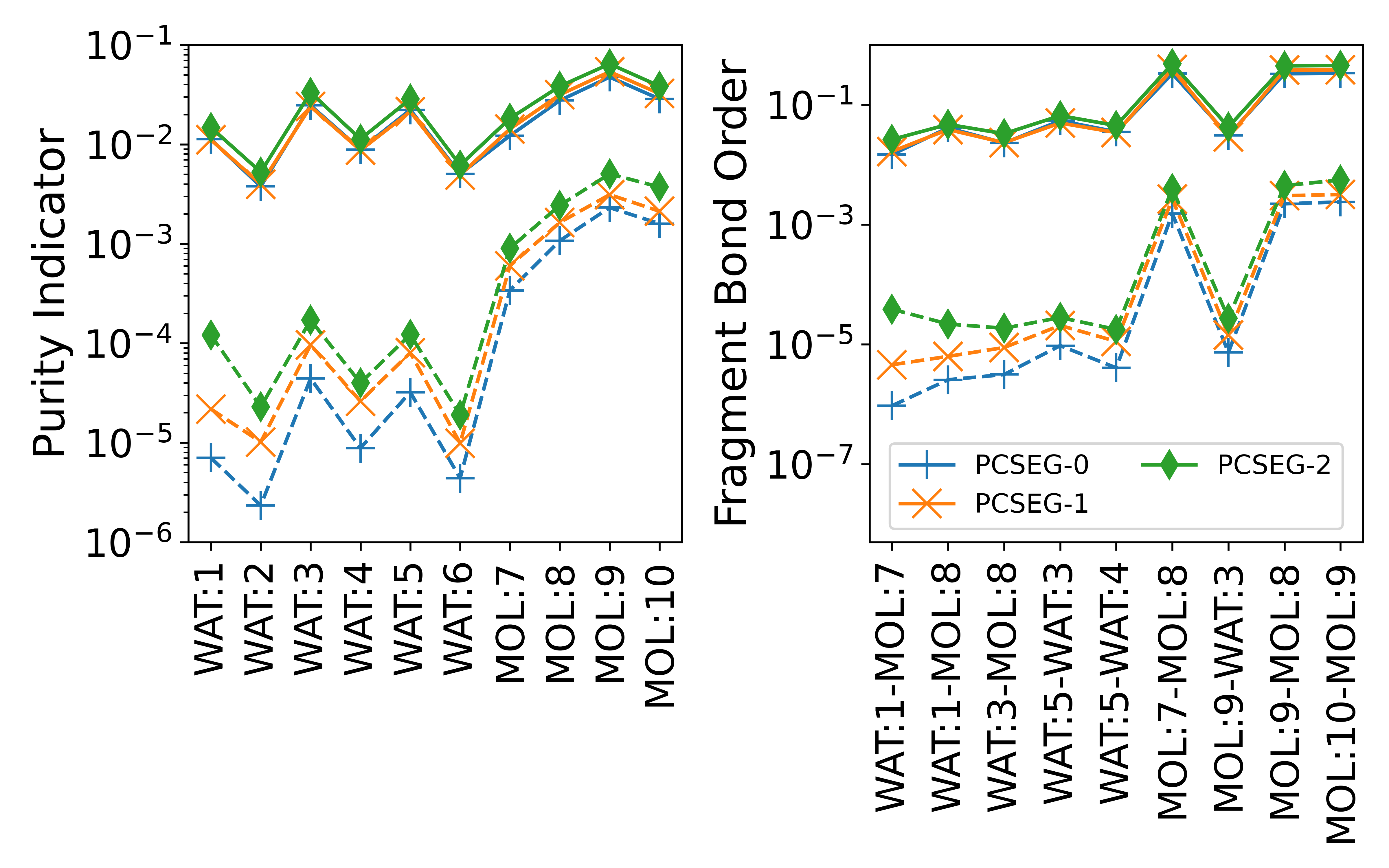}
    \caption{Comparison of (absolute) $\Pi$ values and $B$ for occupied (solid) and core electron only (dashed) density matrices. When the core density matrix is used, we use a value of $2$ for $q_{AA}$, since we only consider second row atoms.}
    \label{fig:core}
\end{figure}

We now recompute $\Pi$ and $B$ using $\ddot{K}^C$, the density matrix associated with only the core orbitals (Figure~\ref{fig:core}). For these calculations we use the same Molnupiravir system, the PBE0 functional, and the PCSEG basis set family. With the PCSEG-1 basis, we observe that only 13.47\% of the matrix elements of $\ddot{K}$ are below $1\times 10^{-5}$ in magnitude. This small system is far from the linear scaling cross over point. However, for $\ddot{K}^C$ that value rises to 82.74\%, which reflects the fact that the core--valence gap is much larger than the HOMO-LUMO gap (246.4 vs. 5.1 eV). We observe a corresponding significant decrease in (absolute) $\Pi$ values for the core states and a corresponding decrease in $B$ measures of interaction. The most significant interactions computed with $\ddot{K}^C$ are the covalent bonds which have interaction strengths about an order of magnitude lower than the non-covalent interactions between Molnupiravir fragments and their nearest water molecules computed using $\ddot{K}$.

\begin{figure}
    \centering
    \includegraphics[width=1\columnwidth]{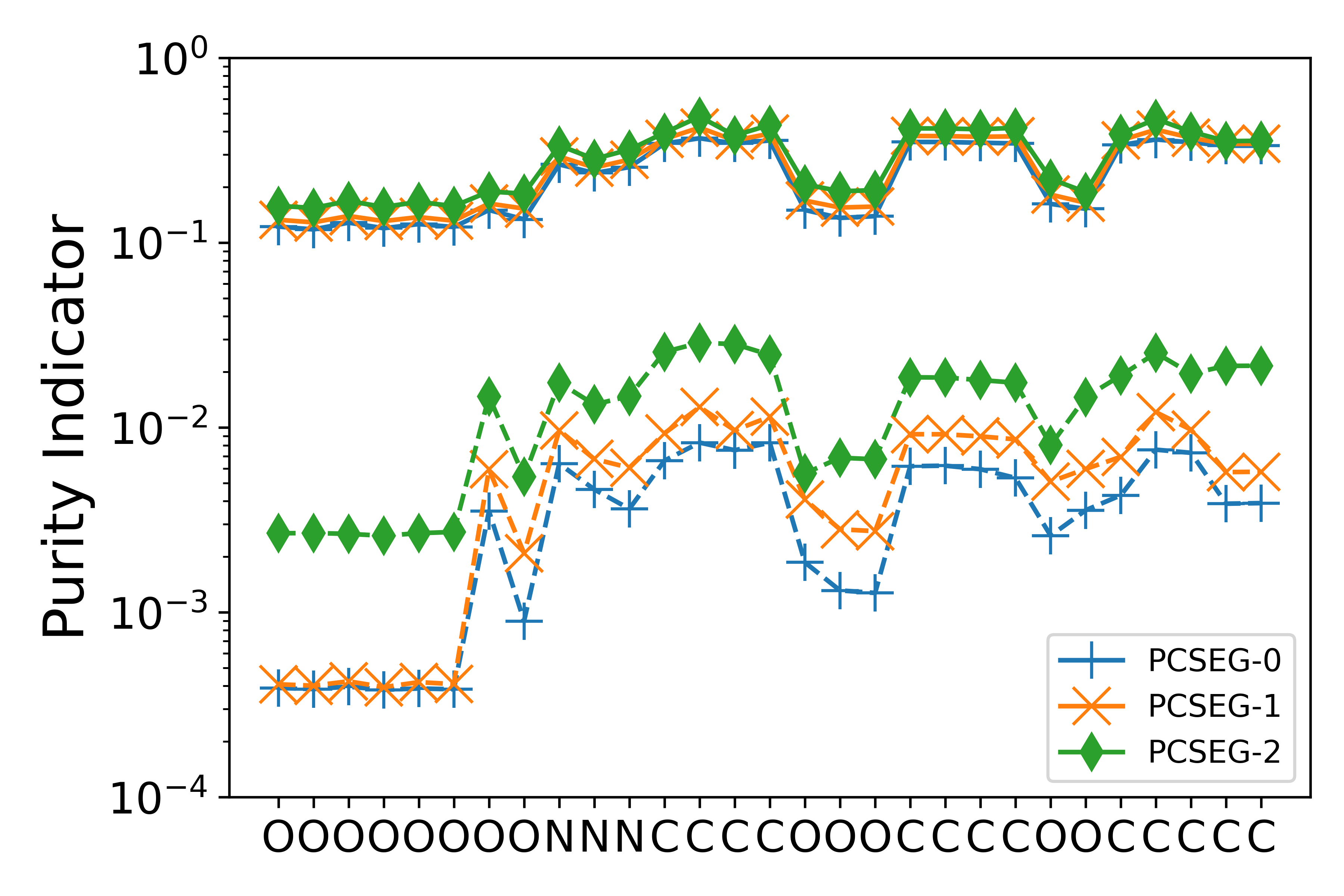}
    \caption{$\Pi$ (absolute) values for a system composed of single atom fragments computed using the occupied (solid) and core electrons only (dashed) density matrices. The six oxygen molecules on the far left are the oxygens from the water molecule fragments.
    }
    \label{fig:pi-core}
\end{figure}

We consider now the $\Pi$ values of the individual atoms of the system (Figure~\ref{fig:pi-core}). If we use $\ddot{K}$, the atomic fragments are clearly not pure, and don't represent a reasonable decomposition of the system. However, if we use $\ddot{K}^C$ we find that the atoms can serve as good fragments, comparable to the water molecules when using $\ddot{K}$. This result can be see as validation of the widely used frozen core approximation. This separation between core states becomes less clear as larger basis sets are used, which will manifest in our analysis of the eigenvalues in the next section.

\section{Eigenvalue Computations}

We now use the insight gained in the previous section to compute the orbital energies of some systems, comparing approaches based on locality in space and energy.

\subsection{Locality in Space}

\begin{figure}
    \centering
    \includegraphics[width=1\columnwidth]{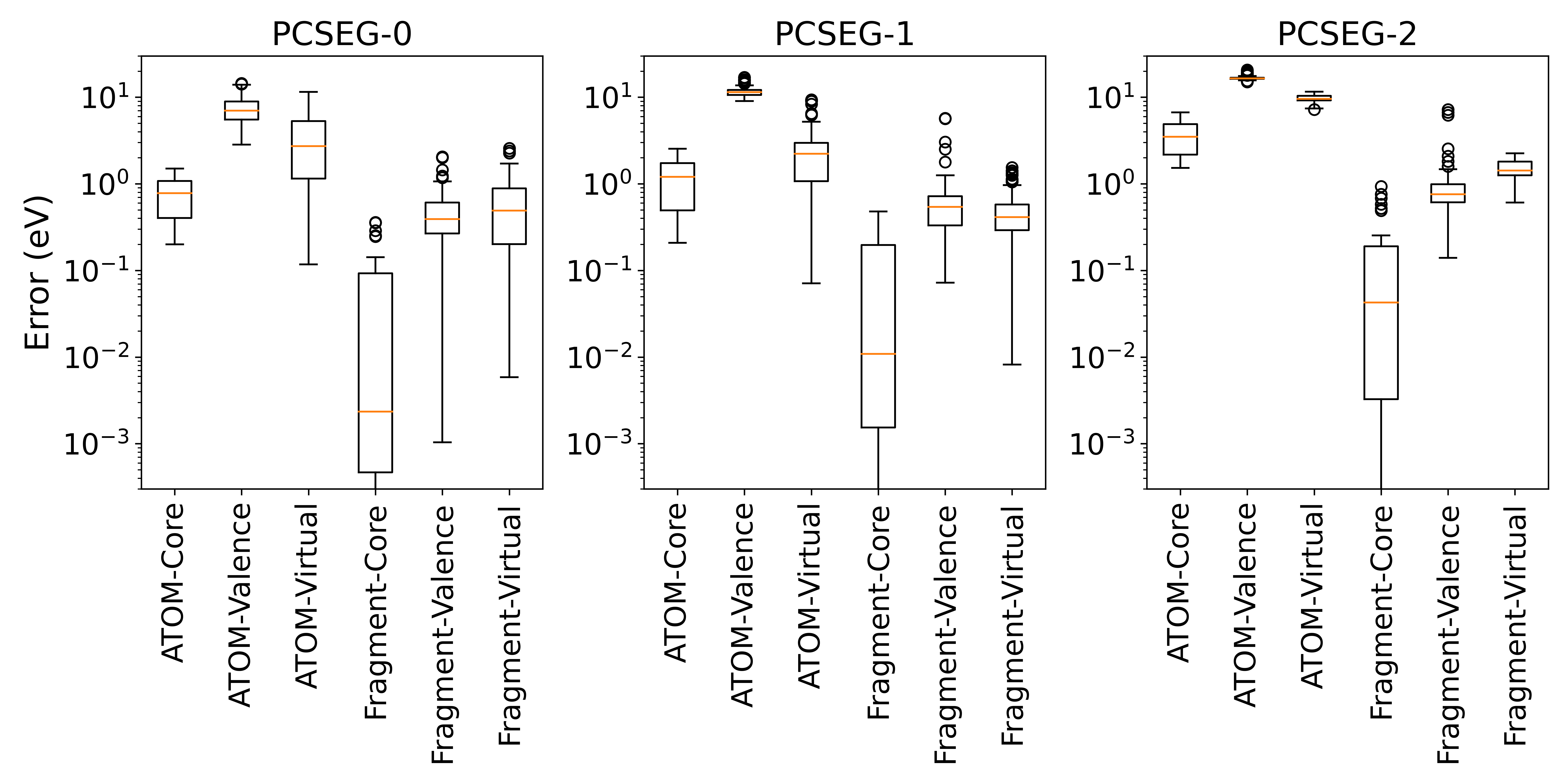}
    \caption{Box plots of the errors in the orbital energies of different parts of the spectrum computed using either the original fragmentation or atomic fragmentation. The whiskers extend the first and third quartile by 1.5 times the interquartile.}
    \label{fig:orb-error}
\end{figure}

Here we attempt to use the purity indicator to construct an approximate block diagonal Hamiltonian. We measure the error of the computed eigenvalues to validate the procedure. We again use the Molnupiravir system with the PBE0 functional. We will analyze the results by studying three areas of the spectrum: the 29 core, 88 valence, and 206 virtual orbitals (chosen for consistency across basis sets). The errors in orbital energies for each of these areas using different fragmentation schemes are plotted in Figure~\ref{fig:orb-error}.

We see a few clear trends from this data. First, the initial fragmentation, which was considered good using BigDFT's $\Pi$ values as a guide, was able to partition the Hamiltonian into block diagonal form, and predict the eigenvalues without much loss of precision. The atomic partitioning was also relatively accurate for the core electrons, similar to the original partitioning's performance on valence and virtual, but was significantly worse for the rest of the spectrum. \hl{This is consistent with the $\Pi$ values computed using $\ddot{K}^C$: the individual atoms are good fragments only for the core orbitals. Conversely, values computed using $\ddot{K}^C$ can't describe a suitable fragmentation for the valence orbitals.} The accuracy for the core electrons also degraded as the basis set was increased, as predicted by Figure~\ref{fig:pi-core}.

\subsection{Embedding Requirements}
\label{subsec:embedreq}

In Table~\ref{tab:nit} we report the energies of the three nitrogen core orbitals. The algorithm presented here uses the fully converged Hamiltonian, as opposed to any fragment method which must approximate both the Hamiltonian and the eigenvalue computation. Clearly a fragment method based on computing isolated atoms would fail to capture nearly any variation in the energies among atoms of the same type. But we also see that even the fairly large fragments of the original fragmentation would almost certainly fail as well; an error of $\sim 0.5$~eV is still too large for practical purposes~\cite{Golze2018}. The cause of the error in the second nitrogen comes from it being on the border of two fragments.

\begin{table}
\begin{tabular}{l l l l}
\hline
\hline
\multicolumn{4}{c}{PCSEG-0} \\
Atomic & -393.63 & -393.63 & -391.05\\
Fragment & -394.42 & -394.33 & -391.83\\
Embed & -394.70 &  -394.33 & -391.83 \\
Full & -394.71 & -394.33 & -391.83\\
\multicolumn{4}{c}{PCSEG-1} \\
Atomic & -392.71 & -392.71 & -390.30\\
Fragment & -393.91 & -393.61 & -391.22\\
Embed & -393.94 & -393.89 & -391.21 \\
Full & -393.97 & -393.91 & -391.22\\
\multicolumn{4}{c}{PCSEG-2} \\
Atomic & -390.53 & -389.64 & -388.68\\
Fragment & -394.03 & -393.23 & -391.29\\
Embed & -394.01 & -393.89 & -391.27 \\
Full & -394.03 & -393.91 & -391.29\\
\hline
\hline
\end{tabular}
\caption{\label{tab:nit} Orbital energies (eV) of nitrogen atoms using different eigenvalue approximations.}
\end{table}

To resolve this issue requires the generation of a new set of overlapping fragments, which can be done using QM-CR. In this scenario, each fragment is computed inside an embedding environment defined by the values of $B$. We use a strict cutoff of $1\times 10^{-4}$ and recompute. For PCSEG-0, this involves including just the covalently bonded atoms. With PCSEG-1, one additional carbon atom from the ring is included. Using PCSEG-2, the nitrogen atom on the border of the original fragmentation is now embedded in an environment defined by 9 atoms (see Supplementary Information Sec.~III for a picture). This environment is extremely accurate in reproducing the eigenvalues of the full system at a reduced cost. Nonetheless, as the orbital energies computed using PCSEG-2 are very similar to those computed with PCSEG-1, the larger environment can be interpreted as a spurious basis set effect.

\subsection{Locality in Energy}

When implemented in exact arithmetic, the eigenvalue algorithm that exploits locality in energy has no error. In practice, we will use NTPoly's thresholding of small values for constructing $\ddot{K}^C$, forming the Cholesky vectors, and for the multiplications performed to reduce the matrix dimension. For this test, we used the fourth order trace resetting method with a convergence threshold of $10$ times the sparsity threshold. We continue using the Molnupiravir system, which is too small for linear scaling calculations, yet has a sparse enough $\ddot{K}^C$ matrix for evaluation of any error in a calculation aimed at core orbitals. 

\begin{table}
\begin{tabular}{l c c c}
\hline\hline
Threshold & \%NNZ & MAX & AVG \\
\hline
\multicolumn{4}{c}{PCSEG-0} \\
$1 \times 10^{-7}$ & 72.29 & $1.2\times10^{-3}$ & $2.0\times10^{-4}$ \\
$1 \times 10^{-6}$ & 46.42 & $2.5\times 10^{-4}$ & $7.5 \times 10^{-5}$ \\
$1 \times 10^{-5}$ & 21.48 & $5.1\times 10^{-3}$ & $2.5\times 10^{-3}$\\
$1 \times 10^{-4}$ & 6.99 & $2.5\times 10^{-1}$ & $6.5\times10^{-2}$ \\
\multicolumn{4}{c}{PCSEG-1} \\
$1 \times 10^{-7}$ & 59.78 & $1.8 \times 10^{-5}$ & $6.9 \times 10^{-6}$ \\
$1 \times 10^{-6}$ & 32.52 & $1.6 \times 10^{-2} $ & $1.9 \times 10^{-3}$ \\
$1 \times 10^{-5}$ & 13.66 & $3.1\times10^{-1}$ & $4.7\times10^{-2}$ \\
$1 \times 10^{-4}$ & 4.41 & $2.1\times10^{0}$ & $6.6\times10^{-1}$ \\
\multicolumn{4}{c}{PCSEG-2} \\
$1 \times 10^{-7}$ & 51.90 & $1.9\times10^{-3}$ & $9.2\times10^{-4}$ \\
$1 \times 10^{-6}$ & 21.98 & $1.5\times10^{-1}$ & $6.3\times10^{-2}$ \\
$1 \times 10^{-5}$ & 7.38 & $4.7\times10^{0}$ & $2.5\times10^{0}$ \\
$1 \times 10^{-4}$ & 5.12 & $2.0\times10^{1}$ & $9.3\times10^{0}$ \\
\hline
\hline
\end{tabular}
\caption{\label{tab:dac} Sparsity of the Cholesky vectors (defined as percentage of non-zeros \%NNZ) and errors in orbital energies (Maximum and Average; unit of eV) computed using the local in energy algorithm.}
\end{table}

In Table~\ref{tab:dac} we plot the error in the core eigenvalues using different sparsity threshold values. Similar to $\ddot{K}^C$, the Cholesky vectors are very sparse when a lower threshold is used. The error at a fixed threshold grows with the size of the basis set which is likely related to the conditioning of the basis. With a threshold value of $1\times10^{-6}$ it is possible to compute core eigenvalues with little loss of precision by exploiting locality in energy. 

\subsection{Protein In Water}

\begin{figure}
    \centering
    \includegraphics[width=1\columnwidth]{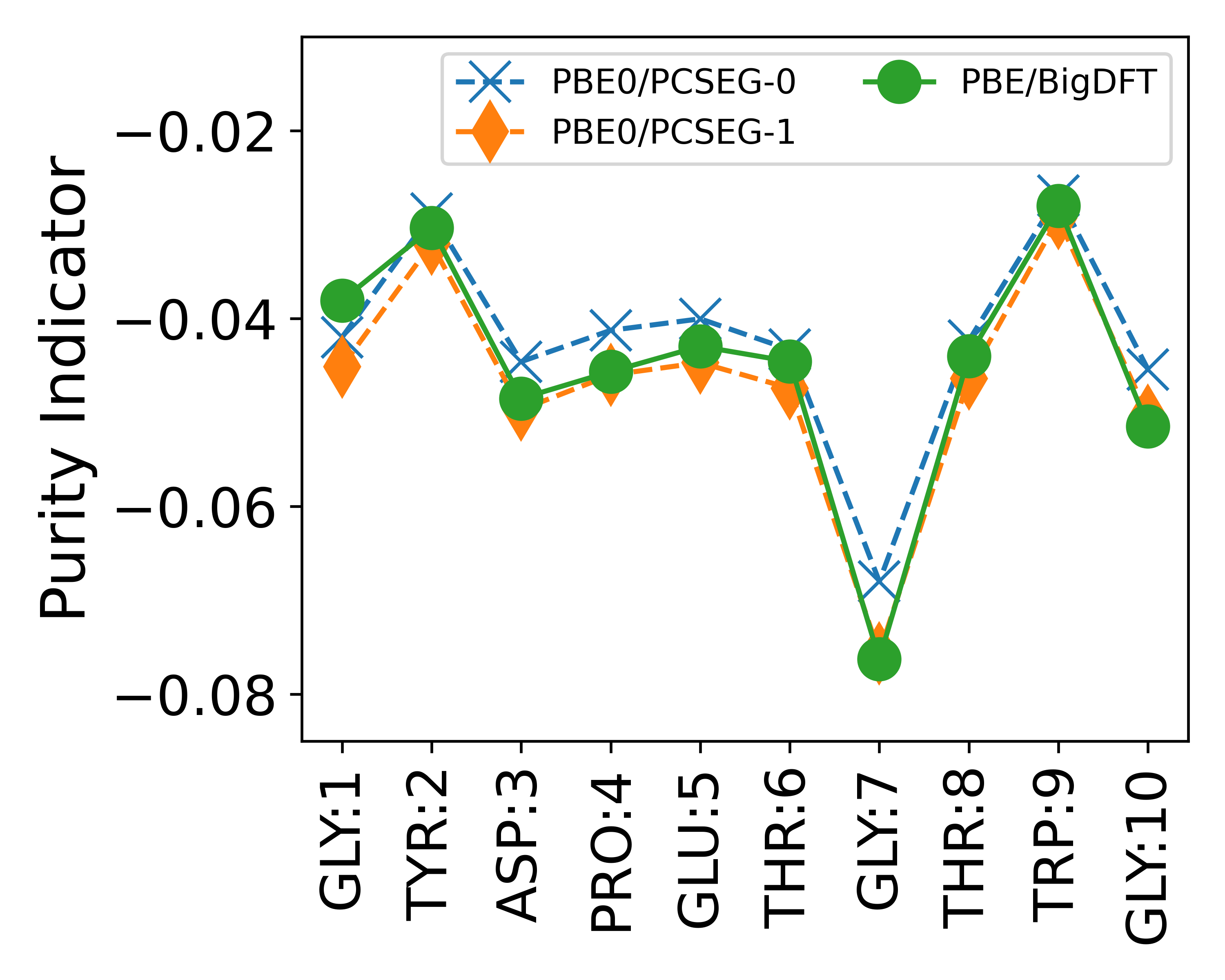}
    \caption{Purity indicator values of the protein amino acids of the 1UAO system computed with BigDFT (PBE) and NTChem (PBE0).}
    \label{fig:amino-acid}
\end{figure}

We will finish by applying our algorithms to larger systems made up of proteins in water. First, we will study the 1UAO protein~\cite{Honda2004} in a sphere of water molecules (3621 total atoms) prepared using CHARMM-GUI~\cite{Jo2008,Jo2014,Lee2016,Lee2020} and minimized with the amber forcefield (see Sec.~\ref{sec:example}). In the original presentation of the purity indicator~\cite{Mohr2017} we proposed a value of $0.05$ as a cutoff for determining the quality of a fragment. This value was picked as an analogy to the p-value used for hypothesis testing in statistics. We now re-evaluate this choice by comparing it to the naturally occurring amino acid fragments of a system. In Figure~\ref{fig:amino-acid}, we plot these purity values as computed with both NTChem and BigDFT. We see a cancellation of effects as the Gaussian orbitals are more diffuse, yet the inclusion of exact exchange leads to a more compact description. As noted in a previous publication~\cite{Dawson2022}, the only residue that falls outside the $0.05$ cutoff is the non-terminal glycine. We thus recommend a loosening of the criteria to $0.08$ for future studies.

The 1UAO protein is made up of just 138 atoms in a large environment ($\sim$10\angstrom{} in radius). It is expected to interact strongly with the solvent environment due to its charged (terminal residues, Asp3, and Glu5) and polar residues (Trp9). It may then be of interest to compute the density of states projected on to the protein, without diagonalizing the entire Hamiltonian, by means of the proposed local in space algorithm. 

In Table~\ref{tab:aa-wat} we plot the error in the core eigenvalues of the protein when computed using the local in space algorithm. Even with no environment the orbital energies are highly accurate. Nonetheless, we can observe some effects from the water molecules. Visualization of the automatically generated environment identifies these water molecules as those participating in hydrogen bonds with the protein (see Supplementary Information Sec.~IV). We also computed three core orbitals using the atomic fragmentation. We choose the atoms with the highest, lowest, and median purity values as tests. A cutoff of $1\times 10^{-4}$ is adequate for the accurate calculation of all three. For the protein system with the PCSEG-1 basis set, that value corresponds to embedding in an environment composed of all nearest neighbor atoms. Based on the results of Sec.~\ref{subsec:embedreq}, this would not be sufficient for triple-$\zeta$ quality basis sets, but can be a simple means of extracting core eigenvalues of calculations at a double-$\zeta$ or lower quality.

\begin{table}
\begin{tabular}{c c c c}
\hline\hline
\multicolumn{4}{c}{Core Full Protein} \\
\hline
$B$ & Environment & AVG & MAX \\
$1 \times 10^{-3}$ & 0 & 0.00499 & 0.05382 \\
$5 \times 10^{-4}$ & 8 & 0.00259 & 0.03489 \\
$1 \times 10^{-4}$ & 22 & 0.00019 & 0.00158 \\
\hline
\multicolumn{4}{c}{Core Single Atoms} \\
\hline
$B$ & Trp9-CG  & Glu5-C & Thr6-OG1 \\
$1 \times 10^{-2}$ & 1.58140 & 1.59768 & 0.42891 \\
$5 \times 10^{-4}$ & 0.01161 & 0.01206 & 0.00708 \\
$1 \times 10^{-5}$ & 0.00124 & 0.00119 & 0.00346 \\
\hline
\hline
\end{tabular}
\caption{\label{tab:aa-wat} (Absolute) Error (eV) in the DoS projected on to the protein when computed with the local in space algorithm at various cutoffs. \hl{Environment size is in number of water molecules when the full protein is treated as a single fragment (Core Full Protein), and the number of environment atoms when each atom is treated as its own fragment (Core Single Atoms).}}
\end{table}

\subsection{Local in Energy Computational Performance}

The local in energy algorithm can be used as a black box solver when the existence of a core-valence gap is known in advance. We thus assess the performance of this algorithm for inclusion in $O(N)$ electronic structure codes using the 1CRN system as a benchmark. We use a cutoff of $1\times10^{-6}$ for filtering small matrix values. With a threshold of $1\times10^{-6}$, $\ddot{K}$ has a sparsity of $5.32\%$, while $\ddot{K}^C$ has a sparsity of $0.38\%$ and the Cholesky vectors $0.80\%$. Being able to dynamically exploit the sparsity that exists in any energy window is one benefit of a matrix element filtering method such as implemented in NTPoly as opposed to the use of a fixed sparsity pattern based on interatomic distances. The largest error in the core eigenvalues is only $0.017$ eV. This error is very similar to what was found for the much smaller Molnupiravir system which suggests the algorithm is robust. 

\begin{table}
\begin{tabular}{c c c c c}
\hline
\hline
Nodes & Full & Purify & Cholesky & Solve \\
\hline
16 & 104.68 & 86.99 & 3.96 & 0.88 \\
32 & 73.14 & 42.99 & 3.16 & 0.88 \\
64 & 47.78 & 28.48 & 3.01 & 0.87 \\
128 & 38.51 & 14.74 & 2.36 & 0.84 \\
256 & 32.59 & 10.93 & 2.64 & 0.98 \\
\hline
\hline
\end{tabular}
\caption{\label{tab:eigtime} Comparison of time to solution (seconds) of full
diagonalization (Full) and the local in energy algorithm \hl{(using the Hamiltonian of the 1CRN in salt water system)}. The local in energy algorithm is composed of three main bottlenecks: density matrix purification (Purify), a pivoted Cholesky factorization (Cholesky), and solving of a small eigenvalue problem (Solve).}
\end{table}

We compare the performance of our local in energy algorithm with the EigenExa solver on Fugaku (Table~\ref{tab:eigtime}). We compare against EigenExa version 2.11 using the one-stage algorithm (\texttt{eigen\textunderscore s}), solving for only the core eigenvalues and no eigenvectors. We used four MPI processes per node and 12 OpenMP threads. Overall, we see that the local in energy algorithm is able to outperform full diagonalization by around a factor of two, particularly when a large number of cores are used. However, the double-$\zeta$ quality basis set remains challenging for density matrix purification, requiring 45 iterations to converge. The pivoted Cholesky calculation represents a strong scaling bottleneck, that likely would benefit from an improved implementation. In practice, for computing the occupied eigenvalues the required projection matrix already available, and the Cholesky and Solve steps are all that is required. We believe there is room for future refinements of this algorithm, and that its benefit will become even stronger for increased system sizes.

\section{Conclusion}

In this paper, we have taken advantage of a newly developed version of the NTChem program to analyze the impact of basis set, density functional, and energy envelope on the measures of our Complexity Reduction Framework. With regards to basis set, we do see the drawback of basis set artifacts on the QM-CR values. However, we conclude that the trends are well preserved across basis sets, enabling the QM-CR analysis to be applied with care. This conservation of trends also existed when studying different functionals. The QM-CR values also show clearly the role of exact exchange in the generation of orbitals with more compact support.

We also explore for the first time, to the best of our knowledge, the intrinsic sparsity of the density matrix associated with only the core orbitals. We found that interactions between core orbitals are significantly reduced compared to valence interactions, a validation of frozen core approximations. The revealed sparsity inspired two new algorithms for the computation of orbital energies of a system. When exploiting locality in space, all calculations can be easily driven from a Python virtual notebook. The only requirement is a code which exposes the matrices in a Python readable format (such as the Matrix Market format implemented by NTPoly) and the assignment of basis functions to atoms. We strongly encourage other code developers to expose a similar interface, which will enable similar investigations in the future. For the local in energy algorithm, a pivoted Cholesky decomposition which operates on sparse matrices is required. Our parallel implementation is freely available in NTPoly, though we hope that more optimized versions will be added to standard solver libraries in the future.

In this study, we only performed calculations on first and second row elements (except sodium and chlorine). For transition metals, the semi-core states should close the core--valence gap significantly. In the future, we hope to use our framework to investigate the locality of these states. We further hope that the insights gained here may lead to new frozen core approximations which can dynamically capture the environments atoms exist in. There appears to be ample opportunity for new algorithmic developments in the area of linear scaling DFT that exploit locality both in space and energy.

\section*{Supplementary Material}

\hl{See the Supplementary Material for a visualization of the 1CRN system, an analysis of the effect of varying the amount of exact-exchange on QM-CR values, and visualizations of the required embedding environments.}

\begin{acknowledgments}
This work was supported by MEXT as ``Program for Promoting Research on the Supercomputer Fugaku'' (Realization of innovative light energy conversion materials utilizing the supercomputer Fugaku, Grant Number JPMXP1020210317). Calculations were also performed using the Hokusai supercomputer system at RIKEN (Project ID: Q22460). LG, NT, and WD acknowledge the joint CEA--RIKEN collaborative action. LER acknowledges an EPSRC Early Career Research Fellowship (EP/P033253/1).

\end{acknowledgments}

\section*{Conflict of Interest}

The authors have no conflicts to disclose.

\section*{Data Availability Statement}

We have made available the main Jupyter notebooks and system geometries online at: \url{https://github.com/william-dawson/CR3-Supplementary}. The remaining data that support the findings of this study are available from the corresponding author upon reasonable request.

\bibliography{main}

\end{document}


\preprint{AIP/123-QED}

\title[Supplementary Information - Complexity Reduction in Density Functional Theory: Locality in Space and Energy]{Supplementary Information - Complexity Reduction in Density Functional Theory: Locality in Space and Energy}
\author{William Dawson}
\email{william.dawson@riken.jp}
\author{Eisuke Kawashima}%
\affiliation{ 
RIKEN Center for Computational Science, Kobe, Hyogo,
650-0047, Japan.
}%
\author{Laura~E.~Ratcliff}
\affiliation{Centre for Computational Chemistry,
School of Chemistry, University of Bristol, Bristol BS8 1TS,
United Kingdom.}
\author{Muneaki Kamiya}
\affiliation{Faculty of Regional Studies, Gifu University, Gifu, 501-1132 Japan}
\author{Luigi Genovese}%
\affiliation{Univ.\ Grenoble Alpes, CEA, IRIG-MEM-L\_Sim, 38000 Grenoble, France.}
\author{Takahito Nakajima}%
\affiliation{ 
RIKEN Center for Computational Science, Kobe, Hyogo,
650-0047, Japan.
}%

\date{\today}

\maketitle

\section{1CRN Protein System}

\begin{figure}[H]
    \centering
    \includegraphics[width=1\columnwidth]{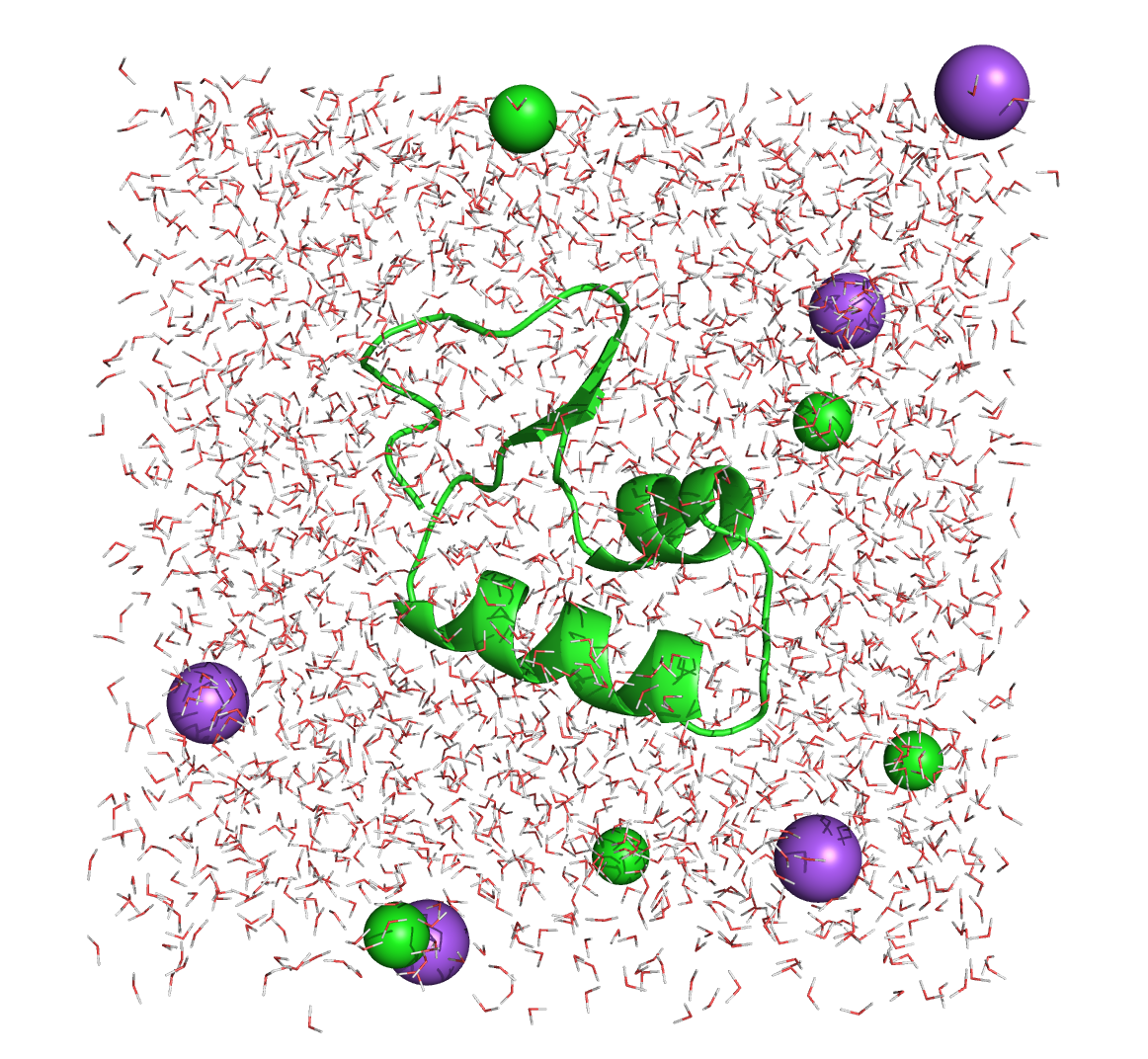}
    \caption{Visualization of the 1CRN in salt water system  (image generated by PyMOL~\cite{PyMOL}). \hl{Sodium (purple) and chlorine (green) ions are represented as spheres.}}
    \label{fig:1crn}
\end{figure}

\section{Effect of Exact Exchange on Purity and Bond Order}

\begin{figure}[H]
\centering
\begin{subfigure}[t]{.5\textwidth}
  \centering
  \includegraphics[width=1\columnwidth]{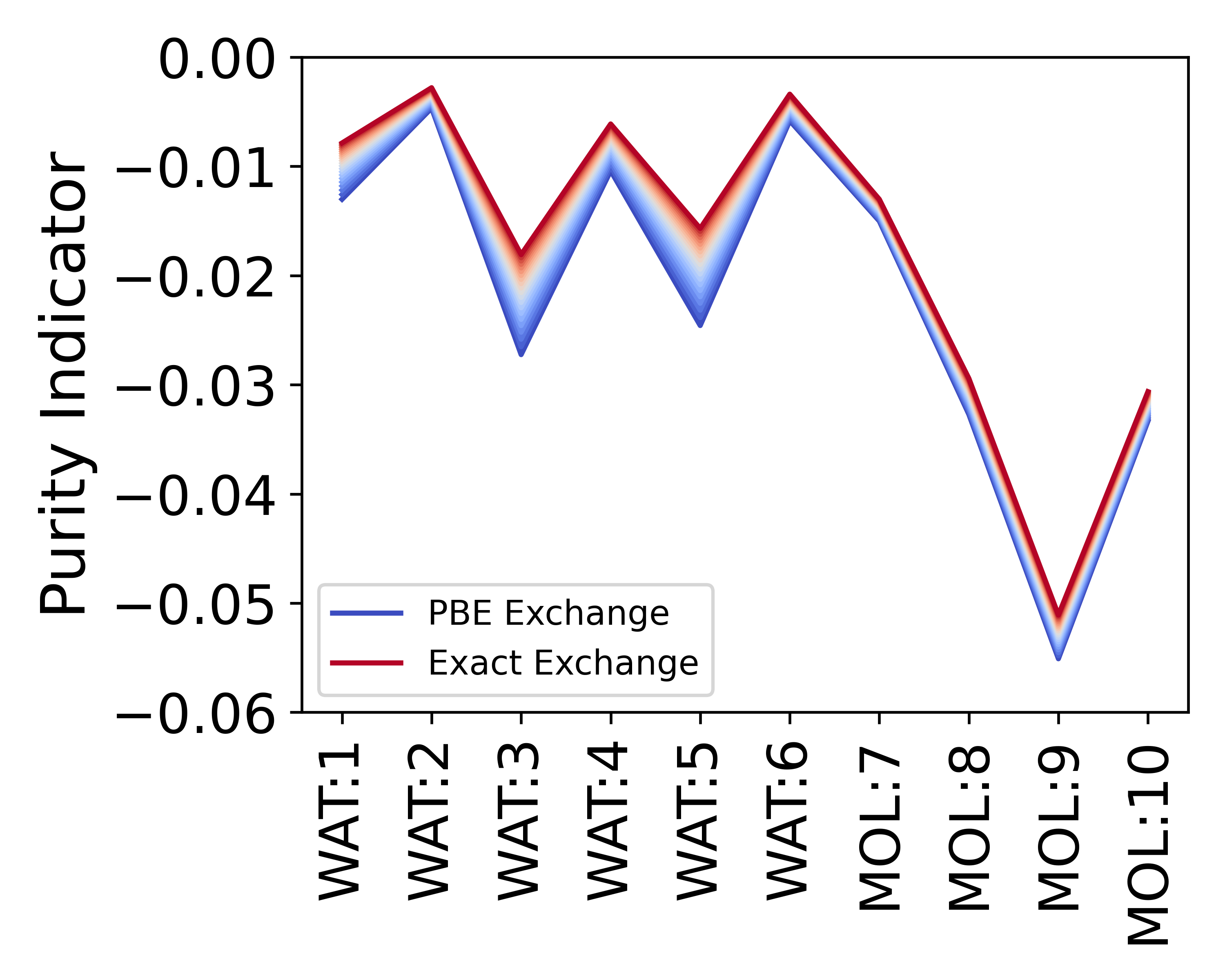}
\end{subfigure}%
\begin{subfigure}[t]{.5\textwidth}
  \centering
  \includegraphics[width=1\columnwidth]{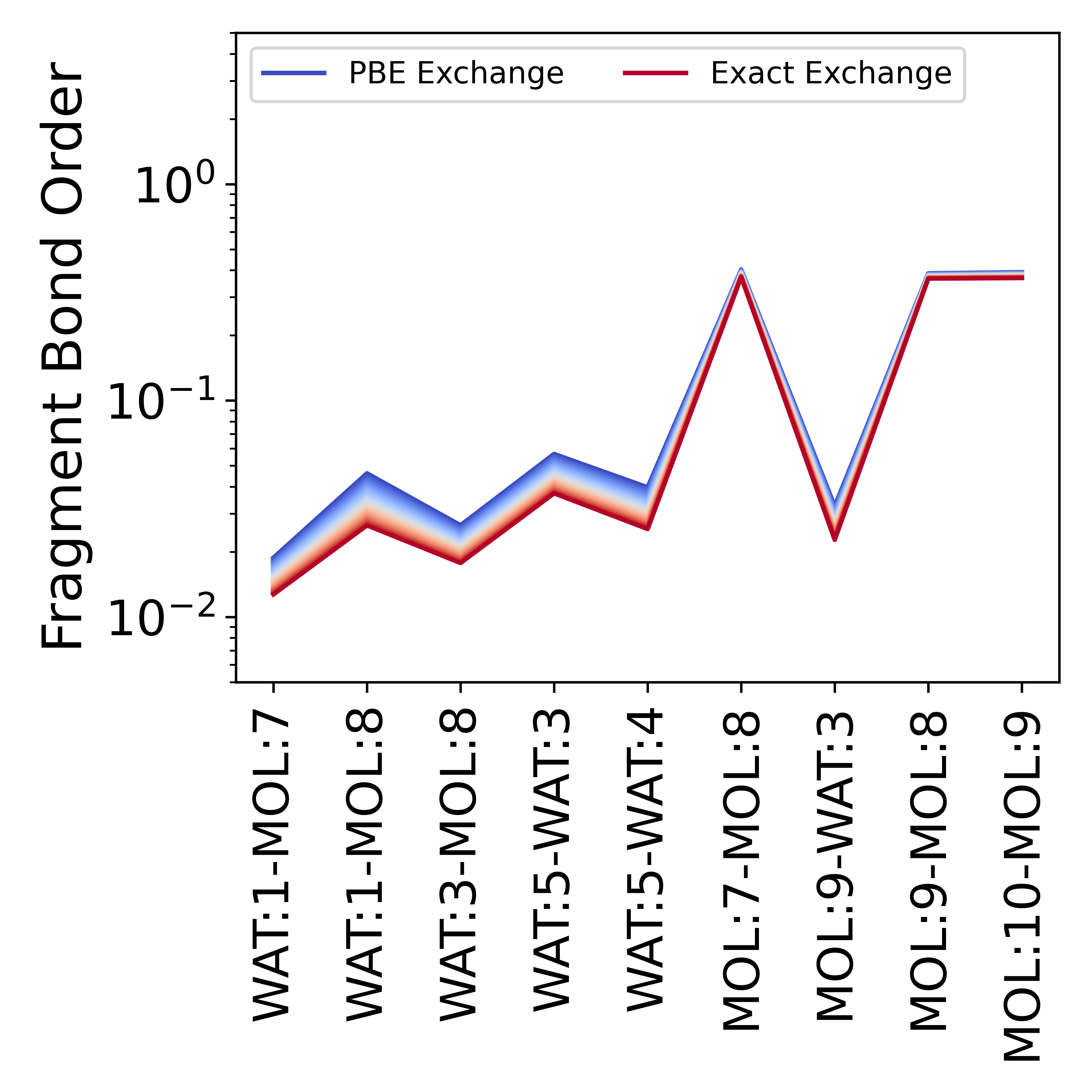}
\end{subfigure}
\caption{Changes in the purity indicator and bond order with the fraction of exact exchange \hl{for the Molnupiravir system using the PCSEG-1 basis set}.}
\label{fig:frac}
\end{figure}

We examine more closely the effect of the fraction of exact exchange on the purity indicator and fragment bond order. We perform calculations using the PBE0 functional~\cite{Adamo1999} with varying levels of exact exchange ranging from $0$ to $1$ in increments of $0.05$. In Figure~\ref{fig:frac} we plot the gradient of values. 

\section{Required Embedding Environment for Border Nitrogen}

The embedding environment for the nitrogen on the border of two fragments as defined by the fragment bond order using a cutoff 
$1 \times 10^{-4}$ and the density matrix from a calculation using the PCSEG-2 basis set~\cite{Jensen2014}.

\begin{figure}[H]
    \centering
    \includegraphics[width=0.5\columnwidth]{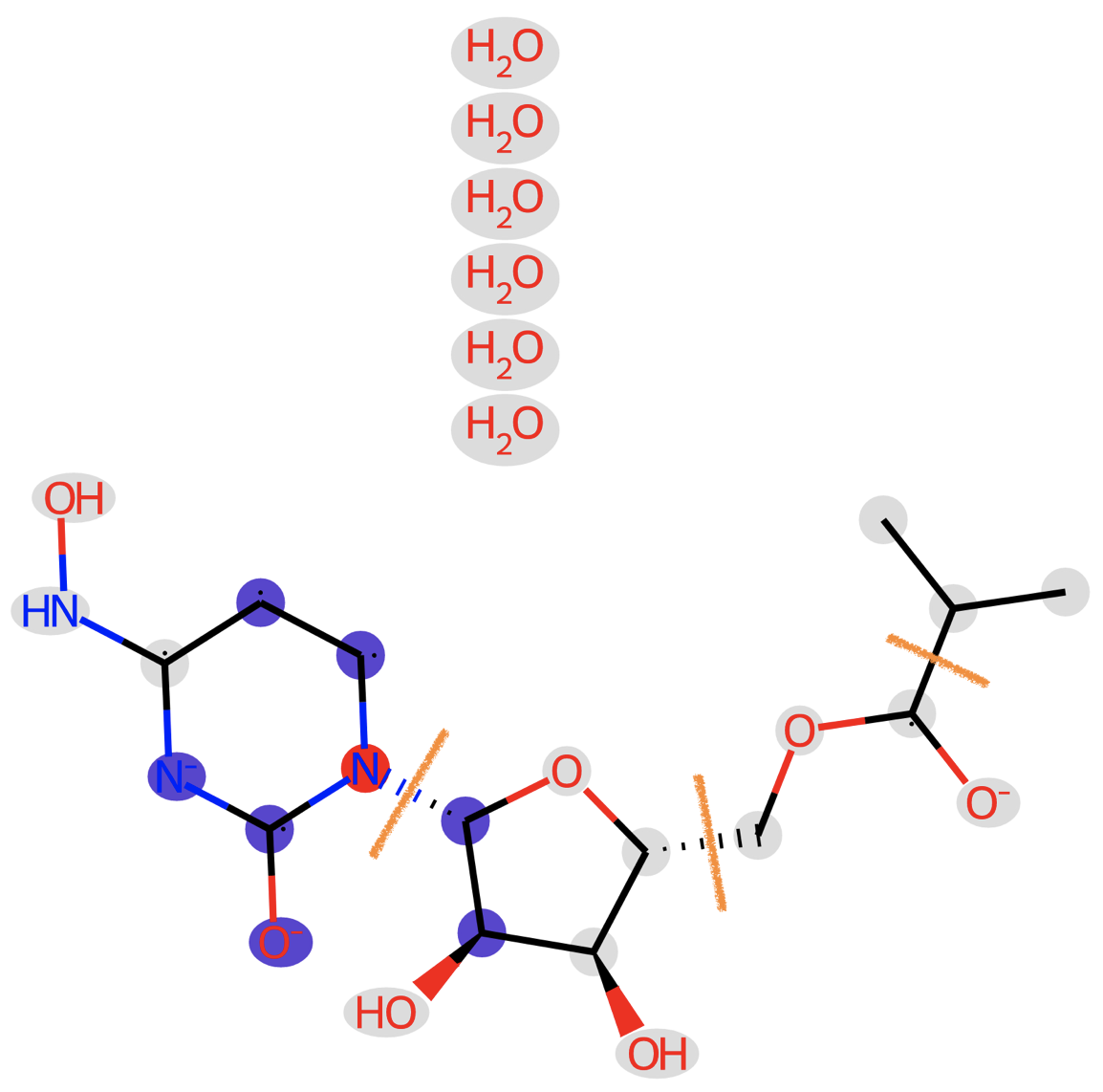}
    \caption{Excluded atoms (gray circles) and required buffer region (blue circles) for reproducing the orbital energy of a given nitrogen atom (red circle) with PCSEG-2. \hl{The borders of the original fragmentation of the Molnupiravir molecule are shown with orange lines.}}
    \label{fig:buffer}
\end{figure}

\section{Protein Environment}

We visualize the molecules included in the embedding environment of the 1UAO protein~\cite{Honda2004} with a cutoff of $B=1\times10^{-4}$ using $\ddot{K}^{C}$.

\begin{figure}[H]
    \centering
    \includegraphics[width=1\columnwidth]{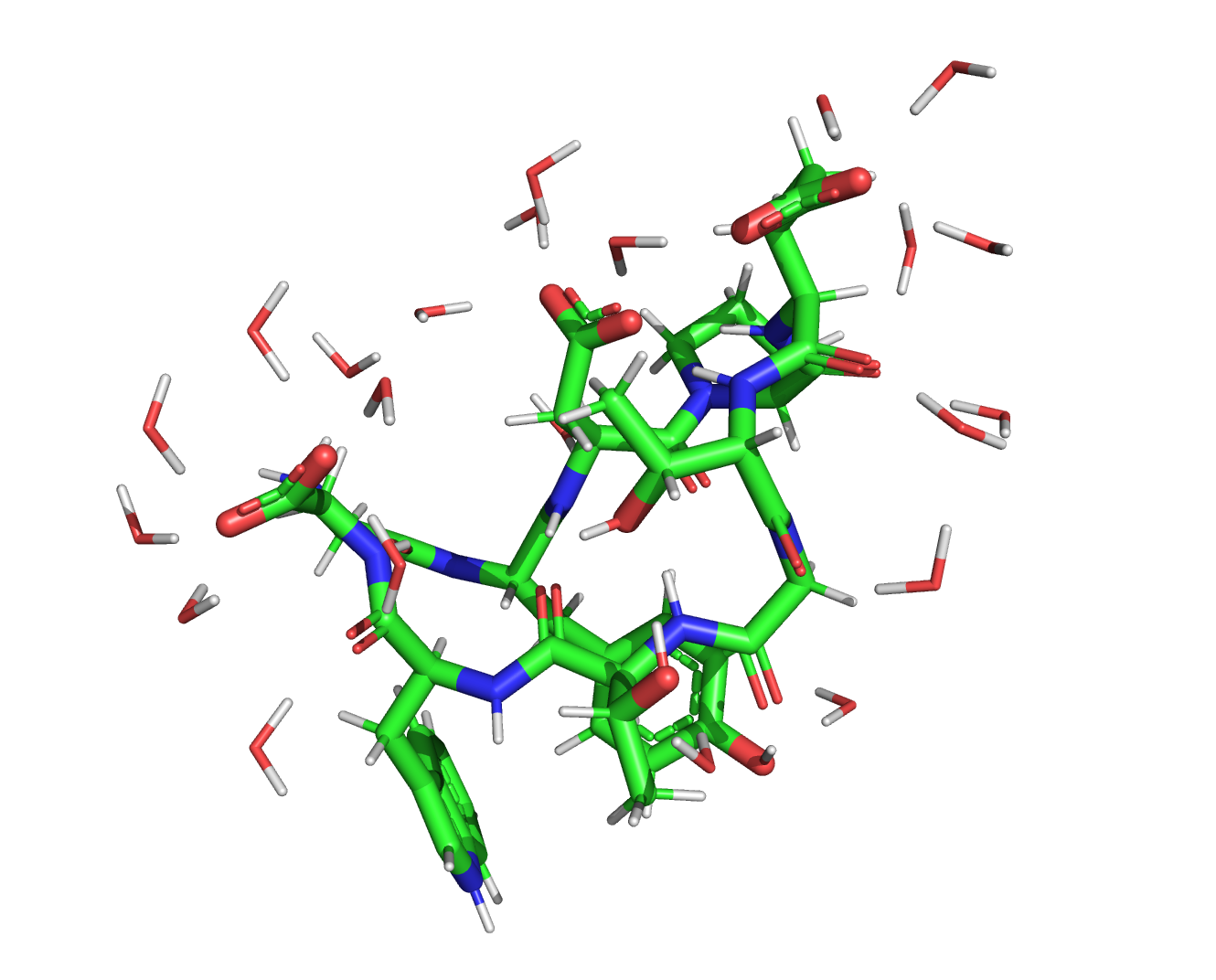}
    \label{fig:hbond}
\end{figure}

\bibliography{supplementary}